\documentclass[aps,pra,twocolumn,showpacs,floatfix]{revtex4}
\usepackage{epsfig}
\usepackage{graphicx}
\usepackage{dcolumn}
\begin{document}
\title{Magic wavelengths for the $np-ns$ transitions in alkali-metal
atoms}
\author{Bindiya Arora}
\affiliation{Department of Physics and Astronomy, University of Delaware, Newark, Delaware 19716-2593}
\author{M. S. Safronova}
\affiliation{Department of Physics and Astronomy, University of Delaware, Newark, Delaware 19716-2593}
\author{Charles W. Clark}
 \affiliation{
Physics Laboratory,
National Institute of Standards and Technology,
Technology Administration,
U.S. Department of Commerce, 
Gaithersburg, Maryland 20899-8410}
\date{\today}
 
\begin{abstract}
Extensive calculations of the electric-dipole matrix elements in
alkali-metal atoms are conducted using the relativistic all-order method. 
This approach is a linearized version of the coupled-cluster method, 
which sums infinite sets of many-body perturbation theory terms. 
All allowed transitions between the lowest $ns, np_{1/2}, np_{3/2}$ states
and a large number of excited states  are considered in these 
calculations and their accuracy is evaluated. 
The resulting electric-dipole matrix elements are used for the 
high-precision calculation of frequency-dependent polarizabilities
of the excited states of alkali-metal atoms. 
We find ``magic'' wavelengths in alkali-metal atoms for which the $ns$
and $np_{1/2}$ and $np_{3/2}$ atomic levels have the same ac Stark shifts, which
facilitates state-insensitive optical cooling and trapping. 
\end{abstract}
\pacs{32.10.Dk, 32.80.Pj, 31.15.Dv, 32.70.Jz}
\maketitle

\section{Introduction}
Recent progress in manipulation of neutral atoms in optical dipole traps offers 
advancement in a wide variety of applications.
One such application is toward the quantum computational scheme, 
which realizes qubits as the internal states of trapped neutral atoms 
\cite{Jaksch}. 
In this scheme, it is essential to precisely localize and control neutral 
atoms with minimum decoherence. 
Other applications include the next generation of atomic clocks, which may attain relative uncertainty
of $10^{-18}$, enabling new tests of fundamental physics, more accurate measurements of fundamental
constants and their time dependence, further 
improvement of Global Positioning System measurements, etc.  

 In a far-detuned optical dipole trap, the potential 
experienced by an atom can be either attractive or repulsive
depending on the sign of the frequency-dependent Stark shift (ac Stark shift) due to the trap light.  
The excited states may experience an ac Stark shift with  an opposite sign
of the ground state Stark shift, affecting the fidelity of the experiments.
A solution to this problem was proposed by Katori 
\textit{et~al.}~\cite{katori}, who suggested that the 
laser can be tuned to a  magic wavelength 
$\lambda_{\rm{magic}}$, where lattice potentials 
of equal depth are produced for the two electronic states 
of the clock transition. 
In their experiment, they demonstrated that a 
$\lambda_{\rm{magic}}$ exists for the $^1S_0 - ^3P_0$ 
clock transition of $^{87}$Sr in an optical lattice. 
Four years later, McKeever \textit{et~al.}~\cite{McKeever} 
demonstrated state-insensitive trapping of Cs atoms at 
$\lambda_{\rm{magic}}$  $\approx$ 935 nm while still 
maintaining a strong coupling for the $6p_{3/2}-6s_{1/2}$ transition.
The ability to trap neutral atoms inside high-Q cavities in the strong coupling regime is 
 of particular importance to the quantum computation and communication schemes ~\cite{McKeever}.

In this paper, we evaluate the magic wavelengths in 
Na, K, Rb, and Cs atoms for which the $ns$ ground state 
and either of the first two $np_j$ excited states experience the same optical 
potential for state-insensitive cooling and trapping. 
We accomplish this by matching the ac polarizabilities 
of the atomic $ns$ and $np_{j}$ states. 
We conduct extensive calculations of the relevant 
electric-dipole matrix elements
using the relativistic all-order method and evaluate 
the uncertainties of the resulting ac polarizabilities. 
We also study the ac Stark shifts of these atoms to determine
the dependence of $\lambda_{\rm{magic}}$ on their hyperfine structure. 

The paper is organized as follows. 
In section~\ref{pol}, we give a short description of the 
method used for the calculation of the ac polarizabilities 
and list our results for scalar and tensor ac polarizabilities. 
In section~\ref{stark}, we discuss the effect of ac Stark 
shifts on the hyperfine structure of the alkali-metal atoms. 
In section~\ref{magic}, we discuss the magic wavelength results 
for each of the atoms considered in this work. 

\section{Dynamic polarizabilities}~\label{pol} 
We begin with an outline of calculations of the ac Stark shift for linearly polarized light, following Refs.~\cite{angel,sch2}. 
Angels and Sandars~\cite{angel}
discussed  the methodology for the calculation of the Stark shift and 
parameterization of the Stark shift in terms of the
 scalar and tensor polarizabilities.  
Stark shifts are obtained as the energy eigenvalues
of the Schr$\ddot{\mbox{o}}$dinger equation with interaction operator $V_I$ given by
	\begin{equation}
	V_I=-\vec{\epsilon} \cdot \vec{d},\label{eq-schro}
	\end{equation}
\noindent	
where $\vec{\epsilon}$ is the applied external 
electric field and $\vec{d}$ is the electric-dipole operator.  
The first-order shift associated with $V_I$ vanishes in alkali-metal atoms. 
Therefore, the Stark shift $\Delta E$ of level $v$ is calculated from the second-order expression
	\begin{equation}
	\Delta E=\sum_k\frac{\left\langle  j_v m_v\left|V_I\right|j_k m_k\right\rangle \left\langle j_k 
m_k\left|V_I\right|j_v 	m_v\right\rangle}{E_v-E_k}\label{eq-shift},
	\end{equation}
\noindent
where the sum over $k$ includes all intermediate states 
allowed by electric-dipole transition selection rules, and    
$E_k$ is the energy of the state $k$. 

Using the Wigner-Eckart theorem, one finds that $\Delta E$ 
can be written as the sum ~\cite{dmitry}
	\begin{equation} \label{eq9}
	\Delta 
E=-\frac{1}{2}\alpha_0(\omega)\epsilon^2-\frac{1}{2}\alpha_2(\omega)\frac{3m_j^2-j_v(j_v+1)}{j_v(2j_v-1)}~\epsilon^2 ,
	\end{equation}
\noindent	
where $\alpha_0(\omega)$ and $\alpha_2(\omega)$ are the scalar 
and tensor ac polarizabilities, respectively, of an atomic state $v$. 
The laser frequency $\omega$  is assumed to be 
several linewidths off-resonance. 
Here, the polarization vector of the light defines the $z$ direction.

The scalar ac polarizability $\alpha_0(\omega)$ of an atom can 
be further separated into an ionic core contribution $\alpha_{\rm{core}}(\omega)$ 
and a valence contribution $\alpha_{0}^v(\omega)$. 
The core contribution has a weak dependence on the frequency for the values of $\omega$ relevant to this work. 
Therefore, we use the static ionic core polarizability values calculated using the random-phase 
approximation (RPA) in Ref.~\cite{datatab2}. 
The valence contribution $\alpha_{0}^v$($\omega$) to the static polarizability of a monovalent atom in a state $v$ is given by~\cite{1}
      \begin{equation}
         \alpha_{0}^v(\omega)=\frac{2}{3(2j_v+1)}\sum_k\frac{{\left\langle 
          k\left\|d\right\|v\right\rangle}^2(E_k-E_v)}{          
         (E_k-E_v)^2-\omega^2}, \label{eq-1}
      \end{equation}
where $\left\langle k\left\|d\right\|v\right\rangle$ is 
the reduced electric-dipole (E1) matrix element. 
The experimental energies $E_i$ of the most important 
states $i$ which contribute to this sum have been compiled
for the alkali atoms in Refs.~\cite{NIST1,NIST, NIST2}.  
Unless stated otherwise, we use atomic units (a.u.) for all 
matrix elements and polarizabilities throughout this paper: the numerical values of the elementary charge, $e$, the reduced Planck constant, $\hbar = h/2 \pi$, and the electron mass, $m_e$, are set equal to 1. 
The atomic unit for polarizability can be converted to SI units via 
$\alpha/h$~[Hz/(V/m)$^2$]=2.48832$\times10^{-8}\alpha$~[a.u.], 
where the conversion coefficient is $4\pi \epsilon_0 a^3_0/h$ 
and the Planck constant $h$ is factored out. 
   
The tensor ac polarizability $\alpha_{2}(\omega)$ is given by  \cite{tensor}:
      \begin{eqnarray}   
         \alpha_{2}(\omega)&=&-4C\sum_k(-1)^{j_v+j_k+1}
         \left\{   
         \begin{array}{ccc}
         j_v & 1 & j_k \\
         1 & j_v & 2 \\
         \end{array}
         \right\} \nonumber \\
        & &\times \frac{{\left\langle 
          k\left\|d\right\|v\right\rangle}^2(E_k-E_v)}{ 
         (E_k-E_v)^2-\omega^2}, \label{eq-2}\\ 
        C &=&  
       \left(\frac{5j_v(2j_v-1)}{6(j_v+1)(2j_v+1)(2j_v+3)}\right)^{1/2}. \nonumber
     \end{eqnarray}     

The ground state ac polarizabilities of alkali-metal atoms 
have been calculated to high precision~\cite{bin1}.
However, no accurate systematic study of the ac polarizabilities 
of the excited states of alkali-metal atoms is currently available. 
The polarizability calculations for the excited $np$ states 
are relatively complicated because in addition to $p-s$ 
transitions they also involve $p-d$ transition matrix elements. 
The matrix elements involving the $nd$ states are generally more difficult 
 to evaluate accurately, especially for the heavier alkalies. 
\begin{table}
\caption{\label{Rbp2}Contributions to the $5p_{3/2}$ scalar ($\alpha_{0}$) and tensor ($\alpha_{2}$) polarizabilities at $\lambda$=790~nm in Rb 
and their uncertainties in units of $a_0^3$. The absolute values of corresponding reduced electric-dipole matrix elements 
$d$ (in $ea_0$) and the corresponding transition wavelength in vacuum $\lambda_{\rm{res}}$ (in nm) are also given.}
\begin{ruledtabular}
\begin{tabular}{lcrrr}
\multicolumn{1}{c}{Contribution} &
\multicolumn{1}{c}{$\lambda_{\rm{res}}$} &
\multicolumn{1}{c}{$d$} &
\multicolumn{1}{c}{$\alpha_0$}&
\multicolumn{1}{c}{$\alpha_2$}\\
\hline
$5p_{3/2}-5s_{1/2}$ & 780& 5.977 &  -4153(5)  & 4153(5)  \\
$5p_{3/2}-6s_{1/2}$ & 1367& 6.047 &   -92(1)   & 92(1)  \\
$5p_{3/2}-7s_{1/2}$ & 741& 1.350 &   41.1(1)& -41.1(1)   \\
$5p_{3/2}-8s_{1/2}$ & 616& 0.708 &   2.88(2)& -2.88(2)  \\
$5p_{3/2}-9s_{1/2}$ & 566& 0.466 &   0.922(6) & -0.922(6)   \\
$5p_{3/2}-10s_{1/2}$& 539& 0.341 &   0.430(3) & -0.430(3)    \\[0.5pc]
$5p_{3/2}-4d_{3/2}$ & 1529& 3.633 &  -26.9(4)  & -21.5(4)    \\
$5p_{3/2}-5d_{3/2}$ & 776& 0.665 &   36(3)    &  29(3)  \\
$5p_{3/2}-6d_{3/2}$ & 630& 0.506 &   1.6(3)   &  1.3(3)   \\
$5p_{3/2}-7d_{3/2}$ & 573& 0.370 &   0.60(9)  &  0.48(7)  \\
$5p_{3/2}-8d_{3/2}$ & 543& 0.283 &   0.30(4)  &  0.24(3)   \\
$5p_{3/2}-9d_{3/2}$ & 526& 0.225 &   0.18(2)  &  0.14(2)   \\[0.5pc]
$5p_{3/2}-4d_{5/2}$ & 1529& 10.899&  -242(4)   &  48.4(8)   \\
$5p_{3/2}-5d_{5/2}$ & 776& 1.983 &   317(28)  & -63(6)   \\
$5p_{3/2}-6d_{5/2}$ & 630& 1.512 &   14(3)    &  -2.9(6)  \\
$5p_{3/2}-7d_{5/2}$ & 573& 1.104 &   5.4(8)   &  -1.1(2)  \\
$5p_{3/2}-8d_{5/2}$ & 543& 0.845 &   2.7(3)   &  -0.54(6)  \\
$5p_{3/2}-9d_{5/2}$ & 526& 0.672 &   1.6(2)   &  -0.31(3)  \\[0.5pc]
$\alpha_{\rm{tail}}$&  &          &      19(14) &    -5(5)       \\
$\alpha_{\rm{core}}$&  &          &      9.1(5) &                 \\
Total               &  &  &  -4060(32)&   4184(9)         
\end{tabular}   
\end{ruledtabular}
\end{table}

In this work, we calculate $np-n^{\prime}d$ transition matrix elements 
using the relativistic all-order method~\cite{CC,relsd} and 
use these values to accurately determine the $np_{1/2}$ and $np_{3/2}$ state ac polarizabilities. 
In the relativistic all-order method, all single and double (SD)
excitations of the Dirac-Fock (DF) wave function are 
included to all orders of perturbation theory.  
For some matrix elements, we found it necessary to also include single, double and partial triple (SDpT) excitations into the wave functions (SDpT method).
We conduct additional semi-empirical scaling of our 
all-order SD and SDpT values  
where we expect scaled values to be more accurate or for more accurate evaluation 
of the uncertainties. 
The scaling procedure has been described in Refs.~\cite{relsd,CC2,1}. 

We start the calculation of the $np$ state valence 
polarizabilities using Eqs. (\ref{eq-1}) 
and (\ref{eq-2}). 
For the wavelength range considered in this work, 
the first few terms in the sums over $k$ give the dominant contributions.
Therefore, we can separate the $np$ state valence polarizability 
into a main part, $\alpha_{\rm{main}}$, that includes these dominant terms, and a remainder,  $\alpha_{\rm{tail}}$.
We use a complete set of DF wave functions on a 
nonlinear grid generated using B-splines~\cite{relsdrb} in all our calculations. 
We use 70 splines of  order 11 for each value of the angular momentum. 
A cavity radius of 220 a.u. is chosen to accommodate all valence 
orbitals of $\alpha_{\rm{main}}$. In our K and Rb calculations, we include all $ns$ states up to $10s$
and all $nd$ states up to $9d$; $11s$, $12s$, and $10d$ are also added for Cs.   
Such a large number of states is needed to reduce uncertainties in the 
remainder $\alpha_{\rm{tail}}$. 
We use the experimental values compiled in Ref.~\cite{volz} 
along with their uncertainties for the first $np-ns$ matrix elements, for example the 
$5p_{j}-5s$ matrix elements in Rb.
 We use the SD scaled values for some of the $np-n^{\prime}d$ and $np-n^{\prime}s$ 
matrix elements in the cases where it was essential to reduce 
the uncertainty of our calculations and where the scaling is expected to produce 
more accurate results based on the type of the dominant correlation corrections. 
This issue is discussed in detail in Refs.~\cite{usca,Safronova:8s07} and references therein.   

In Table~\ref{Rbp2}, we give the contributions 
 to the scalar and tensor 
polarizabilities of the Rb $5p_{3/2}$ state at 790 nm to illustrate the 
details of the calculation.  
The absolute values of the corresponding reduced electric-dipole matrix elements, $d$,  
used in the calculations are also given. The contributions from the main term are 
listed separately. We also list the resonant wavelengths $\lambda_{\rm{res}}$ corresponding to each transition
to illustrate which transitions are close to  790 nm. 
As noted above, we use the experimental 
values for the $5p_{3/2}-5s$ matrix element from the Ref.~\cite{volz}. 
We use the recommended values for the $5p_{3/2}-4d_j$ transitions derived from the 
Stark shift measurements \cite{milhun2}  in Ref.~\cite{KRb}. 
We find that the contribution of the $5p_{3/2}-5s$ transition 
is dominant since the wavelength of this transition 
($\lambda_{\rm{res}}$ = 780 nm) is the closest to the laser wavelength. 
The next dominant contribution for the scalar polarizability 
is from the $5p_{3/2}-5d_{5/2}$ transition ($\lambda_{\rm{res}} = 776$~nm).
While the contribution from this transition is less than one tenth of the 
dominant contributions, it gives the dominant contribution to the final
uncertainty owing to a very large correlation correction to the $5p_{3/2}-5d_{5/2}$ reduced 
electric-dipole matrix element.  In fact, the lowest-order DF value for this transition is only 
0.493~a.u. while our final (SD scaled) value is 1.983~a.u. 
 We take the uncertainty in this transition to be the 
 maximum difference of our final values and \textit{ab initio} SDpT 
 and scaled SDpT values. 
While the $5p_{3/2}-5d_{3/2}$ transition has almost 
the same transition wavelength owing to the very small 
fine-structure splitting of the $5d$ state, 
the corresponding contribution is nine times smaller owing to the fact that the $5p_{3/2}-5d_{3/2}$ 
reduced electric-dipole matrix element is smaller than the 
$5p_{3/2}-5d_{5/2}$ matrix element by a factor of three. 
As expected, the contributions from the core and tail 
terms are very small in comparison with the total polarizability values
at this wavelength. 

\begin{table}
\caption{\label{comp1}Comparison of static polarizabilities of $np_{1/2}$ and $np_{3/2}$ states with other experiments and 
theory. $^a$Ref.~\cite{poldere}, $^b$Ref.~\cite{ekstrom}, $^c$Ref.~\cite{windholzm}, $^d$derived from 
Ref.~\cite{milhun} D1 line Stark shift measurements and recommended values for ground state polarizability from 
Ref.~\cite{pol-andrei}, $^e$Ref.~\cite{krenn}, $^f$Ref.~\cite{win1}, $^g$derived from Ref.~\cite{milhun2} D1 line Stark 
shift measurement and ground state polarizability measurement from Ref.~\cite{amini}, $^h$derived from Ref.~\cite{tanner1} 
D2 line Stark shift measurement and ground state polarizability  from Ref.~\cite{amini}.  Units: a$_0^3$.}
\begin{ruledtabular}
\begin{tabular}{lllll}
\multicolumn{1}{l}{Na } &
\multicolumn{1}{l}{$\alpha_0$(3p$_{1/2}$) } &
\multicolumn{1}{l}{$\alpha_0$(3p$_{3/2}$)} &
\multicolumn{1}{l}{$\alpha_2$(3p$_{3/2}$)}\\
\hline
 Present& 359.9     	 & 361.6        &-88.4		 		\\   
 Other  & 359.7$^a$    & 361.4$^a$    &-88.0$^a$ 	  	\\ 
 Exp.		& 359.2(6)$^b$ & 360.4(7)$^b$ &-88.3(4)$^c$    \\
\hline
K				& $\alpha_0$(4p$_{1/2}$)		& $\alpha_0$(4p$_{3/2}$)   & $\alpha_2$(4p$_{3/2}$) 		\\
\hline		
Present &   602    		  &  613         & -109	  		\\   
Other   &   605$^a$    	&  616$^a$		 & -111$^a$     	\\ 
Exp.		&   606.7(6)$^d$&  614(10)$^e$ &-107(2)$^e$								\\    
\hline   
Rb			& $\alpha_0$(5p$_{1/2}$)		& $\alpha_0$(5p$_{3/2}$)   & $\alpha_2$(5p$_{3/2}$)  \\
\hline		
Present &  805       	  &  867  		   & -167  			\\   
Other   &  807$^a$    	&  870$^a$		 & -171$^a$	    \\ 
Exp.		&  810.6(6)$^d$ &  857(10)$^e$ & -163(3)$^e$					 			\\   
\hline   
Cs			& $\alpha_0$(6p$_{1/2}$)		& $\alpha_0$(6p$_{3/2}$)   & $\alpha_2$(6p$_{3/2}$) \\
\hline		
Present & 1338        	& 1650    	 	 & -261 					\\   
Other   & 1290$^f$    	& 1600$^f$   	 & -233$^f$		    \\ 
Exp.		& 1328.4(6)$^g$ & 1641(2)$^h$  & -262(2)$^h$					   		      			   		      
\end{tabular} 
\end{ruledtabular}
\end{table}

In Table~\ref{comp1}, we compare our results for 
the first excited $np_{1/2}$ and $np_{3/2}$ state 
static  polarizabilities for Na, K, Rb, and Cs with 
the previous experimental and theoretical studies. 
The measurements of the ground
state static polarizability of Na by Ekstrom 
\textit{et al.}~\cite{ekstrom}
were combined  with the experimental Stark shifts from 
Refs.~\cite{windholzm,windholzn} to predict precise 
values for the $3p_{1/2}$ and $3p_{3/2}$ scalar polarizabilities ~\cite{ekstrom}.  
The tensor polarizability of the $3p_{3/2}$ state of 
Na has been measured  by Windholz \textit{et al.}~\cite{windholzm}.
The Stark shift measurements for K and Rb have 
been carried out by Miller \textit{et al.}~\cite{milhun} for D1
lines and by Krenn \textit{et al.}~\cite{krenn} for D2 lines.
We have combined these Stark shift measurements with 
the recommended ground state polarizability values from Ref.~\cite{pol-andrei}
to obtain the $np_{j}$ polarizability values that we quote as
experimental results.
The $np_{3/2}$ tensor polarizabilities in K and Rb were measured in 
Ref.~\cite{krenn}.
Accurate D1 and D2 Stark shift measurements for 
Cs have been reported in Refs.~\cite{milhun2,tanner1}. 
The most accurate experimental measurement  of the $6s$ ground state polarizability 
from Ref.~\cite{amini} has been used to derive  
the values of the $6p_{1/2}$ and $6p_{3/2}$ state polarizabilities in Cs quoted 
in Table~\ref{comp1}. Our results are in excellent agreement with the experimental values.

We note that we use our theoretical values for the $4p-3d$ transitions in K and
  $5p-4d$ transitions in Rb to establish the accuracy of our approach. 
  We use more accurate recommended values  for these transitions derived from the 
  experimental Stark shifts  \cite{milhun2}  in Ref.~\cite{KRb} in all
  other calculations in this work, as described in the discussion of Table~\ref{Rbp2}.

\section{ac Stark effect for hyperfine levels}\label{stark}
In the above discussion, we neglected the 
hyperfine structure of the atomic levels. 
However,  it is essential 
to include the hyperfine structure which is affected by the presence of external electric field
for practical applications discussed in this work. 
In this section, we calculate the eigenvalues of the 
Hamiltonian $H$ representing the combined effect 
of Stark and hyperfine interactions. 
Then, we subtract the hyperfine splitting from the above 
eigenvalues to get the ac Stark shift of a hyperfine level. 
This value is used to calculate the ac Stark shift of 
the transition from a hyperfine level of excited $np$ state 
to a hyperfine level of ground $ns$ state. 

\subsection{Matrix elements of the Stark operator}
First, we evaluate the matrix elements of Stark operator 
in the hyperfine basis. 
The energy difference between two hyperfine levels 
is relatively small for cases considered in this work, and the hyperfine 
levels are expected to mix even if small electric fields are applied. 
Therefore, the Stark operator now  has non-zero off-diagonal matrix elements. 
The general equation for the matrix elements is given by
	\begin{widetext}
	\begin{equation}
	\Delta E_{F,F''}=\sum_k\frac{\left\langle I jFM\left|V_I\right|I_k j_k F_kM_k\right\rangle \left\langle I_k 		
j_k F_kM_k\left|V_I\right|I'' j''F''M''\right\rangle }{E_v-E_k},\label{eq-3}
	\end{equation}
	\end{widetext} 
\noindent	
where  $\left|IjFM\right\rangle$ represents 
atomic states in hyperfine basis, $I$ is the nuclear spin, and 
$\bf{F}=\bf{I}+\bf{j}$.

The interaction operator $V_I$ given by Eq.~(\ref{eq-schro}) does not 
affect the nuclear spin \textit{I}. 
In addition, the shifts due to $V_I$ are not 
high enough to cause mixing between two levels 
with different angular momentum $j$.  
As a result, $\Delta E_{F,F''}$ is diagonal in $\textit{I}$ and $\textit{j}$. 
These approximations enable us to label the states in 
hyperfine basis as $\left|F M\right\rangle$, and Eq. (\ref{eq-3}) can be simplified as

	\begin{equation}
	\Delta E_{F,F''}=\sum_k\frac{\left\langle FM\left|V_I\right|F_kM_k\right\rangle \left\langle  F_kM_k\left|V_I\right| 	
F''M''\right\rangle }{E_v-E_k}.\label{eq-11}
	\end{equation}
	
One can write the above matrix element as
	\begin{equation}
	\Delta E_{F,F''}=\left\langle FM\left|V_{II}\right|F''M''\right\rangle,\label{eq-4}
	\end{equation}	
	
\noindent
where the Stark shift operator $V_{II}$ is 
defined in terms of $\lambda$ operator as
	\begin{eqnarray}
	V_{II}  &=& V_I\lambda V_I  \rm{,}\\
	\lambda &=& \sum_k\frac{\left|F_k M_k\right\rangle \left\langle F_k M_k\right|}{E_v-E_k}\label{offd}.
	\end{eqnarray}
	
If the applied electric field is in the $z$ direction, 
then the energy shifts are diagonal in $M$. 
Thus, the matrix elements can be written as
	\begin{equation}
	\Delta E_{F,F''}=\left\langle FM\left|V_{II}\right|F''M\right\rangle.
	\end{equation}

We use the Wigner-Eckart theorem to carry out
the angular reduction, i.e. sum over the magnetic quantum numbers. 
Then, the matrix elements can be written in terms of scalar 
and tensor polarizabilities as
	\begin{eqnarray}
	\left\langle FM\left|V_{II}\right|F''M\right\rangle&=&-\frac{1}{2}\alpha_0(\omega)\epsilon^2\delta_{F,F''} \nonumber 
\\  &-&\frac{1}{2}\alpha_2(\omega)\epsilon^2\left
\langle 	FM\left|Q\right|F''M\right\rangle.\label{eq-h}
	\end{eqnarray}
\noindent
The first term in the equation above, containing the scalar polarizability, results in  the equal shifts of all of  the 
hyperfine levels and is non-zero only for the 
diagonal matrix elements ($F=F''$). 
The tensor part mixes states of different $F$ through Q operator. 
The non-zero matrix elements of Q are
	\begin{eqnarray}
	&&\left\langle FM\left|Q\right|F''M\right\rangle = \left[\frac{(j+1)(2j+1)(2j+3)}{j(2j-1)}\right]^{1/2}\nonumber \\ 	  
	       &&\times (-1)^{I+j+F-F''-M}\sqrt{(2F+1)(2F''+1))} \nonumber \\
	&&\times   
		\left(    
			\begin{array}{ccc}
			F & 2 & F'' \\
			M & 0 & -M \\
			\end{array}
		\right) 
		\left\{    
			\begin{array}{ccc}
			F & 2 & F'' \\
			j & I & j \\
			\end{array}
		\right\}.
	\end{eqnarray}
For each magnetic sublevel, there is 
a matrix with rows and columns labeled by $F$ and $F''$. 
Therefore, magnetic sublevels with different values of $|M|$ are shifted by a different amount. 
A detailed discussion of this matrix is given  by Schmieder~\cite{sch2}. 

\subsection{Energy eigenvalues}
Since the Stark interactions considered in this 
work are comparable to the hyperfine interactions, we find the combined shift of a 
hyperfine level by diagonalizing the Hamiltonian given by 
  \begin{equation}
  H=V_{\rm{hfs}}+V_{II},\label{eq-0}
  \end{equation}
  \noindent where $V_{\rm{hfs}}$ is the hyperfine interaction operator. 
In the hyperfine basis, $V_{\rm{hfs}}$ is diagonal with 
the following matrix elements~\cite{birks}
	\begin{eqnarray}
	&& \left\langle FM\left|V_{\rm{hfs}}\right|FM\right\rangle=\frac{1}{2}Az+\nonumber \\ 
	&&
	\frac{3z(z+1)-4I(I+1)j(j+1)}{4I(2I-1)2j(2j-1)}~B\label{eq-hfs},
	\end{eqnarray} 
\noindent		
where $z=F(F+1)-I(I+1)-j(j+1)$, and $A$ and $B$ 
are hyperfine-structure constants~\cite{Dline}.
The matrix elements of $H$ which describe the 
combined effect of the Stark interaction $V_{II}$ 
and hyperfine interaction $V_{\rm{hfs}}$ are given by
\begin{eqnarray}
	V_{F,F'';M}&=& \left\langle FM\left|V_{II}\right|F''M\right\rangle \nonumber \\
	&+&\left\langle FM\left|V_{\rm{hfs}}
	(F = F'')\right|F''M\right\rangle.
\end{eqnarray}
Using Eq.~(\ref{eq-h}) and Eq.~(\ref{eq-hfs}), the above matrix elements can be reduced to a more useful form
\begin{eqnarray}
  V_{F,F'';M} &= &-\frac{1}{2}\alpha_0\epsilon^2\delta_{F,F''} \nonumber \\
  & -&\frac{1}{2}\alpha_2\epsilon^2\left\langle 
FM\left|Q\right|F''M\right\rangle \nonumber \\
&+& \left\langle FM\left|V_{\rm{hfs}}(F=F'')\right|F''M\right\rangle. \label{eq-matel}
\end{eqnarray} 

The combined shift of a hyperfine level is evaluated 
by diagonalizing the matrix formed with $V_{F,F'';M}$.
The resulting diagonal matrix element ($\Delta E_{F,F}$) 
corresponds to the shift in a hyperfine level $F$, 
resulting from  two effects: the hfs interaction 
$V_{\rm{hfs}}$ and the Stark effect $V_{II}$. 
Consequently, we should subtract the hyperfine splitting 
from the these shifts to get the ac Stark shift of a level given by
\begin{equation}
\Delta_{nl_{j}FM}  = \Delta E_{F,F}-\left\langle FM\left|V_{\rm{hfs}}(F=F'')\right|F''M\right\rangle.
\end{equation}

The ac Stark shift 
of the transition from an excited state to the 
ground state $\Delta E({n'l'_{j'}F'M'}\rightarrow{nl_{j} FM})$ is determined as the difference 
between the ac Stark shifts of the two states.  
We calculate the magic wavelength where the 
ac Stark shift of the $np-ns$ transition is equal to zero. 
The results of the calculation are presented in the next section.

\section{Magic wavelengths for the $np-ns$ transitions}~\label{magic}

We define the magic wavelength $\lambda_{\rm{magic}}$ 
as the wavelength where the ac polarizabilities 
of the two states are the same, leading to 
zero ac Stark shift for a corresponding transition. 
For $np-ns$ transitions considered in this work, it is found at the crossing of the
ac polarizability curves for the $ns$ and $np$ states. 
In the case of the $np_{3/2}-ns$ transitions, the magic wavelengths
need to be determined separately for the cases with $m_j=\pm 1/2$ 
and $m_j=\pm 3/2$  owing to the presence of 
the tensor contribution to the total polarizability of 
$np_{3/2}$ state.
According to Eq.~(\ref{eq9}), the total polarizability for the $np_{3/2}$ states
is determined as $\alpha=\alpha_0-\alpha_2$ for  $m_j=\pm 1/2$ 
and $\alpha=\alpha_0+\alpha_2$  for $m_j=\pm 3/2$. 
The uncertainties in the values of magic wavelengths are found as the maximum 
differences between the central value and the crossings of the 
$\alpha_{ns} \pm \delta \alpha_{ns}$ and $\alpha_{np} \pm \delta \alpha_{np}$ 
 curves, where the  $\delta \alpha$
are the uncertainties in the corresponding $ns$ and $np$ polarizability values. 

\begin{table}
\caption{\label{tabNa}Magic wavelengths $\lambda_{\rm{magic}}$ above 500~nm for the 
$3p_{1/2}-3s$ and $3p_{3/2}-3s$ transition in Na and the 
corresponding values of polarizabilities at the magic wavelengths. The resonant 
wavelengths $\lambda_{\rm{res}}$ for transitions contributing to the $3p_{j}$ ac polarizabilities and the 
corresponding absolute values of the electric-dipole matrix elements are also listed. The wavelengths (in vacuum) 
are given in nm  and electric-dipole matrix elements and polarizabilities are
given in atomic units.}
\begin{ruledtabular}
\begin{tabular}{lccccc}
\multicolumn{3}{l}{Transition: $3p_{1/2}~-~3s$} &&&\\ [0.2pc]
\hline
\multicolumn{1}{l}{Resonances} &
\multicolumn{1}{c}{$d$}  &
\multicolumn{1}{c}{$\lambda_{\rm{res}}$} &
\multicolumn{1}{c}{} &
\multicolumn{1}{c}{$\lambda_{\rm{magic}}$} &
\multicolumn{1}{c}{$\alpha(\lambda_{\rm{magic}})$} \\
\hline
$3p_{1/2} - 4s$       & 3.576(1) &    1138.46   & &                &  \\
                      &          &              & &      1028.7(2) &  241(1) \\
$3p_{1/2} - 3d_{3/2}$ & 6.791    &     818.55   & &   & \\
                      &          &              & &      615.88(1) &  1909(2)\\
$3p_{1/2} - 5s$       & 0.757    &     615.59   & &   & \\
$3p_{1/2} - 3s$       & 3.525(2) &     589.76   & &   & \\
                      &          &              & &     589.457  & 52760(100) \\
$3p_{1/2} - 4d_{3/2}$ & 1.917    &     568.42   & &   & \\
                      &          &              & &      566.57(1) &  -1956(3) \\
$3p_{1/2} - 6s$       & 0.391    &     515.03   & &    & \\
	                         & 	     &    &&   514.72(1) & -514(1)\\
	                    \hline 
\multicolumn{3}{l}{Transition: $3p_{3/2} m_j~-~3s$} &$|m_j|$&&\\ [0.2pc]
\hline
 $3p_{3/2} - 4s $     &   5.067(1) & 1140.69 &    &     &\\
	               &            &         &1/2& 984.8(1) &252(1)\\ 
$3p_{3/2} - 3d_{5/2}$ &   9.122(1) &  819.71 &    &     &\\
$3p_{3/2} - 3d_{3/2}$ &   3.041    &  819.70 &    &      &\\
                      &            &         &1/2 &   616.712(1)& 1854(2)\\
$3p_{3/2} - 5s$       &    1.071   &   616.25&    &       &\\
                      &            &         &1/2 &    589.636  &-66230(80)\\  
                      &            &         &3/2 &    589.557(1) &-42(2)\\  
$3p_{3/2} - 3s$       &    4.984(3)&   589.16&    &       &\\
$3p_{3/2} - 4d_{3/2}$ &    0.857   &   568.98&    &      &\\
$3p_{3/2} - 4d_{5/2}$ &    2.571   &   568.98&    &      &\\
	           	&           &      	& 1/2	&  567.43(1)& -2038(3) \\
	           	&           &      	& 3/2	&   566.79(1)& -1976(3)\\
$3p_{3/2} - 6s$        &    0.553   &   515.48     & &   &\\
		        &            &          &1/2 &   515.01(1)& -517(1)\\
\end{tabular}   
\end{ruledtabular}
\end{table}

We also study $\lambda_{\rm{magic}}$ for transitions between particular
$np_{3/2} F'M'$  and $ns FM$ hyperfine sublevels. 
The ac Stark shifts of the hyperfine sublevels of an atomic state are calculated using the method described in the 
previous section. 
In alkali-metal atoms, all magnetic sublevels have 
to be considered separately; therefore, a  
$\lambda_{\rm{magic}}$ is different for the
$np\mbox{ }F'M'-ns\mbox{ }F M$ transitions. 
We include several examples of such calculations. 

We calculated the $\lambda_{\rm{magic}}$ values for $np_{1/2}-ns$ 
and $np_{3/2}-ns$ transitions for all alkali 
atoms from Na to Cs. As a general rule, 
we do not  list the magic wavelengths which are extremely close to 
the resonances. Below, we discuss the calculation of the 
magic wavelengths separately for each atom.  
The figures are presented only for $np_{3/2}$ states 
as they are of more experimental relevance.	
All wavelengths are given in vacuum. 

\begin{figure*}[htb]
  \includegraphics[scale=0.70]{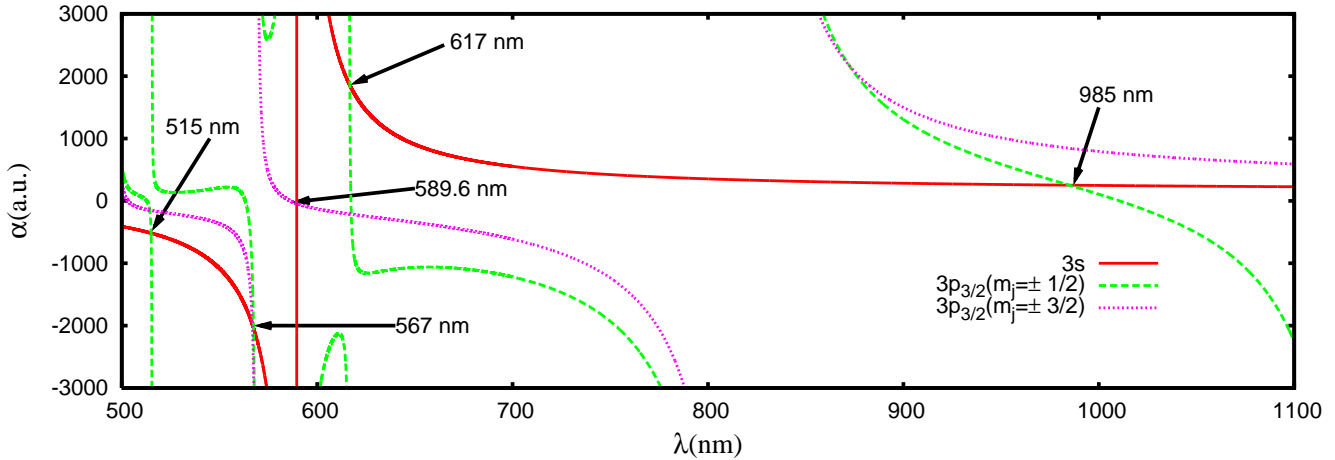}
  \caption{Frequency-dependent polarizabilities of Na atom in the ground and $3p_{3/2}$ states. The arrows
show the magic wavelengths.} 
  \label{figna-1}     
	\end{figure*}
	
		\begin{figure}[htb]
  \includegraphics[scale=0.68]{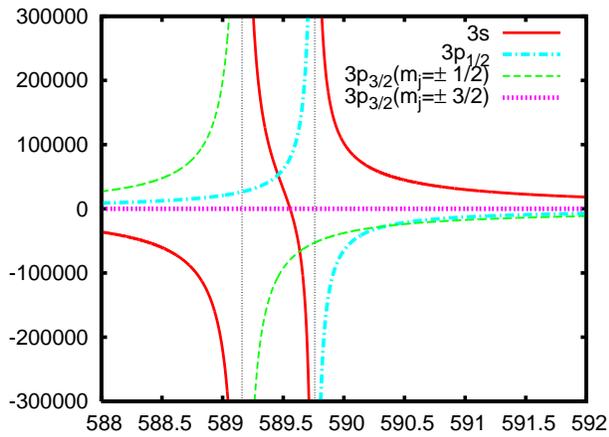}
  \caption{Magic wavelengths for the $3p_{1/2}-3s$ and $3p_{3/2}-3s$ transition of Na.} 
  \label{figna-2}     
	\end{figure}  

	\begin{figure}[htb]
  \includegraphics[scale=0.68]{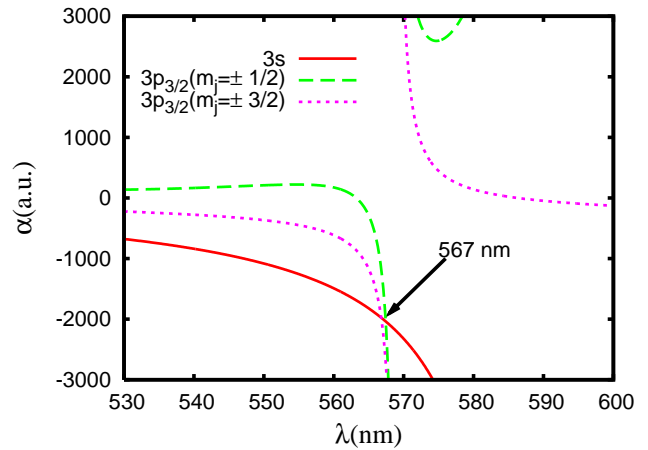}
  \caption{Magic wavelength for $3p_{3/2}-3s$ transition of Na.} 
  \label{figna-3}     
	\end{figure}  
	\begin{figure}[htb]
  \includegraphics[scale=0.68]{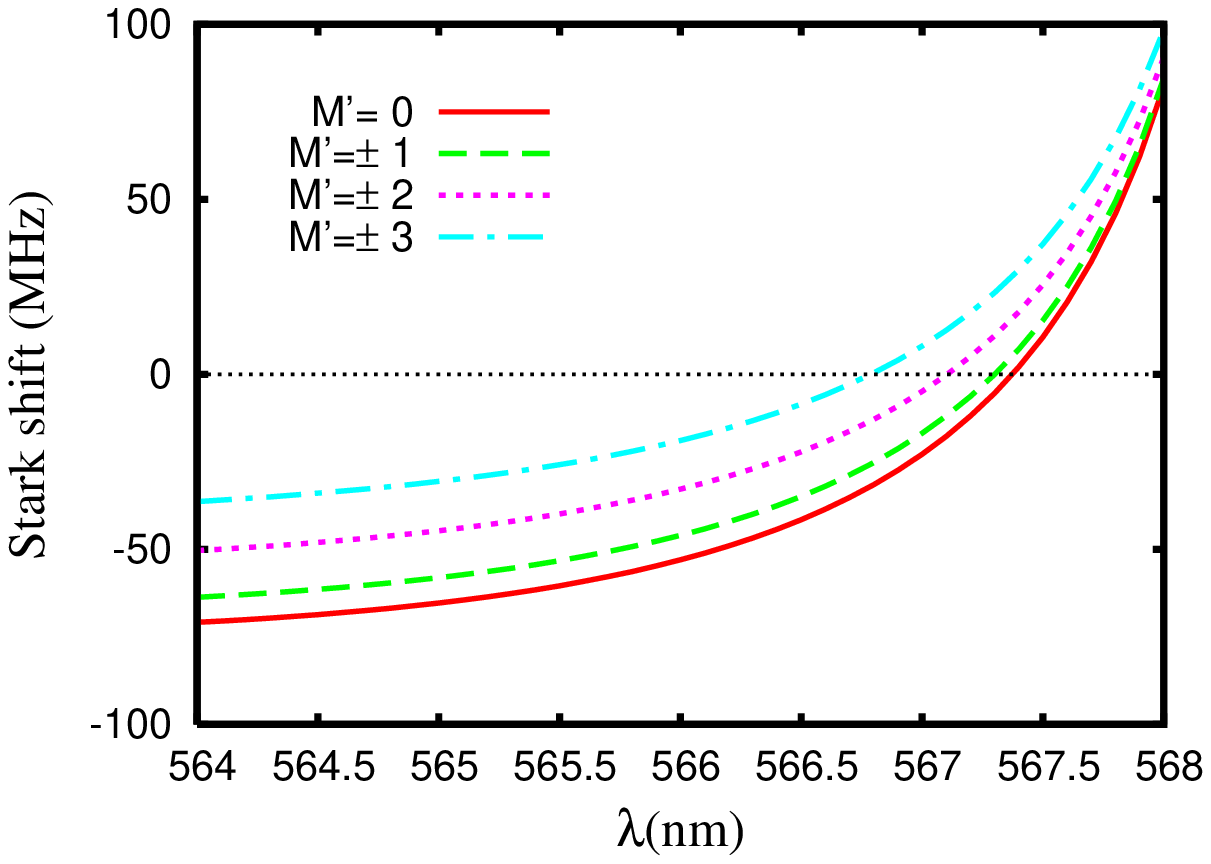}
  \caption{
  ac Stark shifts for the transition from $3p_{3/2} F'=3 M'$ sub levels to $3s FM$ sub levels in Na as a
function of wavelength. The  electric field intensity is taken to be 1 MW/cm$^2$.
} 
  \label{figna-4}   
	\end{figure}
\subsection{Na}

 We list the magic wavelengths $\lambda_{\rm{magic}}$ above 500 nm
  for the $3p_{1/2}-3s$ and $3p_{3/2}-3s$ transitions in Na  and the 
corresponding values of polarizabilities at the magic wavelengths in Table \ref{tabNa}.
  For convenience of presentation, we also list 
  the resonant wavelengths $\lambda_{\rm{res}}$ for transitions contributing 
  to the $3p_{1/2}$ and $3p_{3/2}$ ac polarizabilities and the 
corresponding values of the electric-dipole matrix elements along with their uncertainties. 
 Only two transitions contributing to the ground state polarizabilities are above 500 nm, 
 $3p_{1/2} - 3s$ and $3p_{3/2} - 3s$. Therefore, there is no need to separately list 
 resonant contributions to the ground state polarizability. 
 To indicate the placement of the magic wavelength, we order
 the lists of the resonant and magic wavelengths to indicate
 their respective placement.   The polarizabilities and their uncertainties are calculated as 
 described in Section~\ref{pol}. The transitions up to $np-3s$ and $3p-nd$, $n=6$
 are included into the main term and the remainder is evaluated in the DF approximation. 
 The values of the $3p-3s$ matrix elements are taken from \cite{volz}, the remaining 
 matrix elements are either SD or SD scaled values. The uncertainties in the values of the 
 Na matrix elements were estimated to be 
 generally very small.  The resonant wavelength values are obtained from energy levels from National Institute of Standards and Technology (NIST) database \cite{NIST1}. 
 We assume no uncertainties in the energy values for all elements. 
 
 Since the $3s$ polarizability has only two resonant transitions at wavelengths greater than 500 nm, it is generally
  small except in close vicinity to those resonances. 
   Since the polarizability of the $3p_{1/2}$ state has several contributions 
 from the resonant transitions in this range, it is generally expected that it  crosses the $3s$ polarizability in 
 between of the each pair of resonances listed in Table~\ref{tabNa} unless the wavelength is 
 close to $3p-3s$ resonances. The same is expected in the case of the $3p_{3/2}$ polarizability
  for the $|m_j|=1/2$ cases as described by  Eq.~(\ref{eq9})  ($\alpha=\alpha_0-\alpha_2$). 
  However, when $\alpha=\alpha_0+\alpha_2$ ($m_j= \pm 3/2$ in Eq.~\ref{eq9}) all $3p_{3/2}-ns$
   transitions do not contribute to the total polarizability owing to the exact cancellation of the 
 scalar and tensor contributions for $v=3p_{3/2}$ and $k=ns$ in Eqs.~(\ref{eq-1}) and (\ref{eq-2}).
  In this case, the angular factor in Eq.~(\ref{eq-2}) is exactly $-2/3(2j_v+1)$ leading to exact cancellation 
 of such terms, and the total polarizability comes from the remaining $3p-nd$ contributions which do not cancel out.
  As a result, there are no resonances for the $m_j=\pm 3/2$  cases at the wavelengths
 corresponding to $3p_{3/2}-ns$ transitions leading to substantial reduction in the number of magic wavelengths.
We note that there is a magic wavelength at the 589.557~nm owing to the resonances in the ground state polarizability. 
 The corresponding 
 polarizability value is very small making this case of limited practical use. 
 While there has to be a magic wavelength between $3p_{3/2}-4d_{3/2}$ and $3p_{3/2}-4d_{5/2}$ resonances
 at 568.98~nm, we are not listing it owing to a very small value of the $4d$ fine-structure splitting. 
 We illustrate the magic wavelengths for the $3p_{3/2}-3s$ transition in Fig.~\ref{figna-1}
 where we plot the values of the ac polarizabilities for the ground and $3p_{3/2}$ states. 
 
 It is interesting to consider in more detail 
 the region close to the $3p_{j}-3s$ 
 resonances since in this case one magic wavelength is missing for both $3p_{1/2}-3s$ and 
 $3p_{3/2}-3s$ $\alpha_0-\alpha_2$ cases, one on the side of each $3p-3s$ resonance
 as evident from Table~\ref{tabNa}. 
 We plot the ac polarizabilities for the $3s$, $3p_{1/2}$, and $3p_{3/2} m_j$ states in
  this region in Fig.~\ref{figna-2}. The placements of the $3p_{1/2}-3s$ and $3p_{3/2}-3s$
  resonances are shown by vertical lines. 
  In the case of the $3p_{1/2}$  state, the $3p_{1/2}-3s$ resonance contributes to both 
 ground state and $3p_{1/2}$ polarizabilities. As a result, both of these polarizabilities are 
 large but have opposite sign right of the $3p_{1/2}-3s$ resonance at 589.76~nm
 leading to missing magic wavelength for $3p_{1/2}-3s$ transition between the  $3p_{1/2} - 5s$  and 
$3p_{1/2} - 3s$ resonances. In the $3p_{3/2}-3s$ $\alpha_0-\alpha_2$ case,
there is a missing magic wavelength to the left of the $3p_{3/2}-3s$ 589.12~nm resonance for the same reason.
The values of the $\alpha_0+\alpha_2$ for the $3p_{3/2}$ state are very small and negative 
in that entire region owing to the cancellations of the $3p-3s$ contributions in the 
scalar and tensor $3p_{3/2}$ polarizabilities described above. 

In summary, there is only one case for Na in the considered range of the wavelengths 
  where the magic wavelength exists for all sublevels (567nm)  at close values of the polarizabilities
  (-2000 a.u.) The ac polarizabilities for the $3s$  and $3p_{3/2}$ states  near this magic wavelength 
  are plotted in Fig.~\ref{figna-3}.   
The plot of the
 ac Stark shift for the transition between the hyperfine sublevels near 567~nm is shown in Fig. \ref{figna-4}.
The $\lambda_{\rm{magic}}$ is found at the point where the ac Stark 
shift of the transition from $3p_{3/2}F'=3 M'$ sub levels  
to  $3s FM$ sub levels crosses zero.  
This crossing of ac Stark shift curve occurs close to 567~nm 
which is close to the wavelength predicted by the crossing of polarizabilities
illustrated by Fig.~\ref{figna-3}, as expected.

\begin{table}
\caption{\label{tabK}
Magic wavelengths $\lambda_{\rm{magic}}$ above 600~nm for the 
$4p_{1/2}-4s$ and $4p_{3/2}-4s$ transition in K and the 
corresponding values of polarizabilities at the magic wavelengths. The resonant 
wavelengths $\lambda_{\rm{res}}$ for transitions contributing to the $4p_{j}$ ac polarizabilities and the 
corresponding absolute values of the electric-dipole matrix elements are also listed. The wavelengths (in vacuum) 
are given in nm  and electric-dipole matrix elements and polarizabilities are
given in atomic units.
}
\begin{ruledtabular}
\begin{tabular}{lccccc}
\multicolumn{3}{l}{Transition: $4p_{1/2}~-~4s$} &&&\\ [0.2pc]
\hline
\multicolumn{1}{l}{Resonances} &
\multicolumn{1}{c}{$d$}  &
\multicolumn{1}{c}{$\lambda_{\rm{res}}$} &
\multicolumn{1}{c}{$|m_j|$} &
\multicolumn{1}{c}{$\lambda_{\rm{magic}}$} &
\multicolumn{1}{c}{$\alpha(\lambda_{\rm{magic}})$} \\
\hline
$4p_{1/2} - 5s      $& 3.885(19) &  1243.57&     &           & \\
                   &  	&	       	 & 1/2 & 1227.7(2) & 472(1) \\ 
$4p_{1/2} - 3d_{3/2}$& 7.984(35) &  1169.34&     &&\\
$4p_{1/2} - 4s      $& 4.102(5)  &   770.11&     &&\\
                   &           &         & 1/2 & 768.413(4)& 20990(80)\\
$4p_{1/2} - 4d_{3/2}$& 0.097(57) &   693.82&     &&\\
$4p_{1/2}-  6s      $& 0.903     &   691.30&    &&\\
                   &   	&	        & 1/2 &  690.15(1)& -1186(2)\\
\hline
\multicolumn{3}{l}{Transition: $4p_{3/2} m_j~-~4s$} &$|m_j|$&&\\ [0.2pc]

\hline
$4p_{3/2} - 5s  $& 5.535(26)  & 1252.56  &      &           &\\
        &	&		  &  1/2 &1227.7(2)  &	 472(1)\\
$4p_{3/2} - 3d_{5/2}  $& 10.741(47) & 1177.61  &      &&\\
$4p_{3/2} - 3d_{3/2} $& 3.580(16)  & 1177.29  &      &&\\
        &	&		  & 1/2  &769.432(2) &	 -27190(60)\\
        &            &          & 3/2  & 768.980(3)&  -356(8)\\
$4p_{3/2} - 4s  $&  5.800(8)  &   766.70 &      &&\\
$4p_{3/2} - 4d_{5/2}  $&  0.10(15)  &   696.66 &      &&\\
$4p_{3/2} - 4d_{3/2} $&  0.033(47) &   696.61 &      &&\\
$4p_{3/2} - 6s  $&  1.279     &   694.07 &      &&\\
        &	&		   & 1/2 & 692.32(2) &	  -1226(3)\\
\end{tabular}   
\end{ruledtabular}
\end{table}

\begin{table}
\caption{\label{tabRbCs1}
Magic wavelengths $\lambda_{\rm{magic}}$ above 600~nm for the 
$5p_{1/2}-5s$  transition in Rb and $6p_{1/2}-6s$ 
transitions in Cs  and the 
corresponding values of polarizabilities at the magic wavelengths. The resonant 
wavelengths $\lambda_{\rm{res}}$ for transitions contributing to the $np_{j}$ ac polarizabilities and the 
corresponding absolute values of the electric-dipole matrix elements are also listed. The wavelengths (in vacuum) 
are given in nm  and electric-dipole matrix elements and polarizabilities are
given in atomic units.
}
\begin{ruledtabular}
\begin{tabular}{llcccc}
\multicolumn{1}{l}{Rb} &
\multicolumn{3}{l}{Transition: $5p_{1/2}~-~5s$} &\\ [0.2pc]
\hline
\multicolumn{1}{l}{} &
\multicolumn{1}{l}{Resonances} &
\multicolumn{1}{c}{$d$}  &
\multicolumn{1}{c}{$\lambda_{\rm{res}}$} &
\multicolumn{1}{c}{$\lambda_{\rm{magic}}$} &
\multicolumn{1}{c}{$\alpha(\lambda_{\rm{magic}})$} \\
\hline
& $5p_{1/2} - 4d_{3/2}$&  8.051(67)&   1475.65  &            &\\
&           &    	& 	     &  1350.9(5) &   476(1)\\
& $5p_{1/2} -  6s $&  4.146(27)&   1323.88  &&\\
& $5p_{1/2} -  5s $&  4.231(3) &    794.98  &&\\
&           &    	& 	     &  787.6(1)  &  5417(25)\\
& $5p_{1/2} -  5d_{3/2}$&   1.35(7) &     762.10 &&\\
&           &     	& 	     &  761.5(1)  & -5230(30)\\
& $5p_{1/2} -  7s $&   0.953(2)&     728.20 &&\\
&           &     	& 	     &  727.35(1) &  -1877(3)\\
& $5p_{1/2} -  6d_{3/2}$&   1.07(11)&     620.80 &&\\
 & 	   & 		& 	     &  617.7(7)  &   -490(3)\\
& $5p_{1/2} -  8s $& 0.502(2)  &   607.24   &&\\
&  	  &           &            &	 606.2(1)  &	   -444(1)\\ [0.2pc]
\hline
\multicolumn{1}{l}{Cs} &
\multicolumn{3}{l}{Transition: $6p_{1/2}~-~6s$} &\\ [0.2pc]
\hline
&$6p_{1/2} -  5d_{3/2}$& 7.016(24) &3011.15  &           &\\
&			&         &	       &  1520(3)  &  583(2)\\
&$6p_{1/2} -  7s      $& 4.236(21) &1359.20  & &\\
&$6p_{1/2} -  6s      $&  4.489(7) & 894.59  &    &\\
&$6p_{1/2} -  6d_{3/2}$&  4.25(11) & 876.38  &    &\\
&$6p_{1/2} -  8s      $&  1.026    & 761.10  &&\\
&                      &           &         & 759.40(3) &-1282(3)\\
&$6p_{1/2} -  7d_{3/2}$&  2.05(2)  & 672.51  &  &\\
&                    	&           &         & 660.1(6)  & -513(3)\\
&$6p_{1/2} -  9s      $&  0.548    & 635.63  & &\\
&                    	&           &         &  634.3(2) &  -424(2)\\
&$6p_{1/2} -  8d_{3/2}$&  1.30(2)  & 601.22  & &\\
\end{tabular}   
\end{ruledtabular}
\end{table}

\subsection{K}	

The magic wavelengths $\lambda_{\rm{magic}}$ above 600 nm
  for the $4p_{1/2}-4s$ and $4p_{3/2}-4s$ transitions in K are listed  in Table~\ref{tabK}. 
   Table~\ref{tabK} is structured in exactly the same way as Table~\ref{tabNa}.
   The electric-dipole matrix elements for the $4p-4s$  transitions are taken from \cite{volz}, and the 
   electric-dipole matrix elements for the $4p-3d$ are the recommended values from Ref.~\cite{KRb}
   derived from the accurate Stark shift measurements \cite{milhun}. The resonant wavelengths are obtained from the 
   energy levels compiled in the NIST database \cite{NIST1}. The transitions up to $4p-10s$
   and $4p-9d$
 are included into the main term of the polarizability, and the remainder is evaluated in the DF approximation. 
 In the case of some higher states, such as $9s$, we did not evaluate the uncertainties of the 
 matrix elements where we expect them to be small (below 0.5\%). 
 As a result, the uncertainties in the values of the magic wavelengths near these transitions 
 do not include these contributions and may be slightly larger than estimated.
  In the test case of Rb, the uncertainties are evaluated for all transitions with resonant 
  wavelengths above 600~nm and no significant differences in the uncertainties of the relevant magic
  wavelengths with other elements are observed. 
  
   The main difference between the Na and K calculation is extremely large correlation correction 
   to the values of the $4p-4d$ transitions. The correlation correction  nearly exactly cancels
   the lowest-order DF value leading to a value that is essentially zero within the accuracy of this calculation. 
   As a result, we do not quote the values for the magic wavelength between $4p-4d$ and $4p-6s$
 resonances. We note that these two resonances are very closely spaced (1.5~nm), thus probably making the use of such a magic wavelength impractical.
   Our present calculation places 
  the magic wavelength for the $4p_{1/2}-4s$ transition in the direct vicinity(within 0.01~nm) of the 693.82~nm
  resonance.  We note that the measurement of the ac Stark 
  shift (or the ratio of the $4s$ to $4p$ Stark shifts) near the $4p-4d$ resonance may provide an excellent benchmark
   test of atomic theory. This problem of the cancellation of the lowest and higher-order terms for
    the $np-nd$ transitions
   is unique to K. In the case of Rb, the correlation for the 
  similar $5p-5d$ transition is very large but adds coherently to the DF values. As a result, we were able to evaluate
  the corresponding Rb $5p-5d$ matrix elements with 4.5\% accuracy. The accuracy is further improved for the $6p-6d$
  transitions in Cs.  
  
We also located the magic wavelengths for the $4p_{3/2}-4s$ transition between $4p_{3/2}-3d_{3/2}$
and $4p_{3/2}-3d_{5/2}$ resonances, but found that $m_j=\pm 1/2$ curve crosses the $4s$ polarizability 
very close (within 0.002~nm) to the resonance. Therefore, we do not list this crossing in Table~\ref{tabK}.
We note that $m_j=\pm 3/2$ curve crosses the $4s$ polarizability curve further away from resonance at 
1177.35~nm. The polarizability values for both of these crossings is 500~a.u.
	
	\begin{figure}[!htb]
  \includegraphics[scale=0.68]{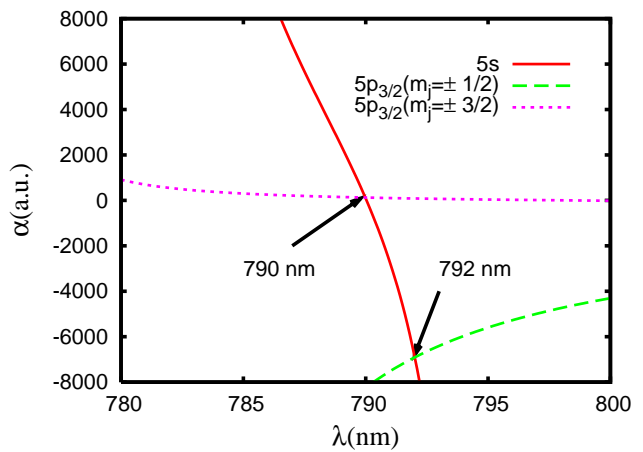}
  \caption{Magic wavelengths for the $5p_{3/2}-5s$ transition of Rb.}
  \label{fig6}     
	\end{figure}
	\begin{figure}[!htb]
  \includegraphics[scale=0.68]{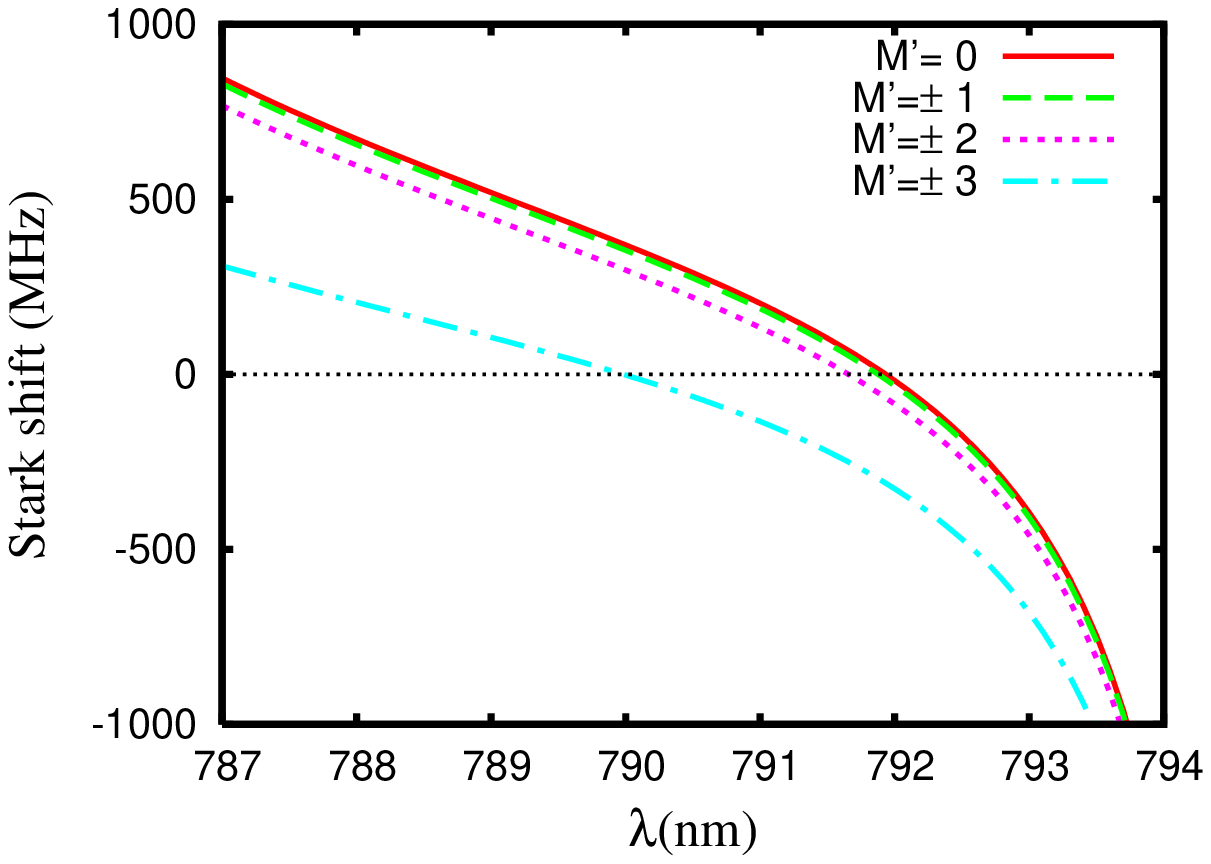}
  \caption{
    ac Stark shifts for the transition from $5p_{3/2} F'=3 M'$ sub levels to $5s FM$ sub levels in Rb as a
function of wavelength. The  electric field intensity is taken to be 1 MW/cm$^2$.
} 
  \label{fig7}     
	\end{figure}\begin{table*}
\caption{\label{tabRbCs3}
Magic wavelengths $\lambda_{\rm{magic}}$ above 600~nm for the 
$5p_{3/2}-5s$  transition in Rb and $6p_{3/2}-6s$ 
transitions in Cs  and the 
corresponding values of polarizabilities at the magic wavelengths. The resonant 
wavelengths $\lambda_{\rm{res}}$ for transitions contributing to the $np_{j}$ ac polarizabilities and the 
corresponding absolute values of the electric-dipole matrix elements are also listed. The wavelengths (in vacuum) 
are given in nm  and electric-dipole matrix elements and polarizabilities are
given in atomic units.}
\begin{ruledtabular}
\begin{tabular}{llcccccc}
\multicolumn{1}{l}{Rb} &
\multicolumn{3}{l}{Transition: $5p_{3/2} m_j~-~5s$} & \multicolumn{2}{c}{$|m_j|=1/2$} &
\multicolumn{2}{c}{$|m_j|=3/2$} \\ [0.2pc]
\hline
\multicolumn{1}{l}{} &
\multicolumn{1}{l}{Resonances} &
\multicolumn{1}{c}{$d$}  &
\multicolumn{1}{c}{$\lambda_{\rm{res}}$} &
\multicolumn{1}{c}{$\lambda_{\rm{magic}}$} &
\multicolumn{1}{c}{$\alpha(\lambda_{\rm{magic}})$}&
\multicolumn{1}{c}{$\lambda_{\rm{magic}}$} &
\multicolumn{1}{c}{$\alpha(\lambda_{\rm{magic}})$} \\
\hline
& $5p_{3/2} -  4d_{5/2} $ &   10.90(9)  &  1529.37  &  	   & 	    &            &\\
& $5p_{3/2} -    4d_{3/2}$ &     3.63(3) &   1529.26&            &        & &\\
& 	  & 		 &           &  1414.8(5)&  456(1)& &\\
& $5p_{3/2} -    6s $ &    6.05(3)  &  1366.87 &            &        & &\\
&           &    	  & 	     & 	 792.00(1)&  -6910(30)& 789.98(2) & 125(35)\\
& $5p_{3/2} -    5s $ &    5.977(4) &   780.24 & & & &\\
& $5p_{3/2} -    5d_{3/2}$ &     0.67(3) &   776.16 & & & &\\
& $5p_{3/2} -    5d_{5/2} $ &    1.98(9)  &   775.98 & & & &\\
&           &    	  & 	     & 	 775.84(1)&  -19990(70)& 775.77(3) & -19700(130)\\
& $5p_{3/2} -    7s $ &    1.350(2) &   741.02 &            & & &\\
& 	 & 	         &          &   740.07(1) &  -2494(4) & &\\
& $5p_{3/2} -    6d_{3/2}$ &    0.51(5)  &   630.10 & & & &\\
& $5p_{3/2} - 6d_{5/2}    $ &   1.51(15)  &   630.01 &   &  &  \\
&                        	&	      	    &        &   627.3(5)&  -533(4)& 626.2(9)&  -528(5)\\
&$5p_{3/2} - 8s$  &   0.708(2)  &   616.13 &  & & &\\
           		    &  	         & &          &614.7(1)&  -477(1)&&\\ [0.2pc]
           		    \hline
\multicolumn{1}{l}{Cs} &
\multicolumn{3}{l}{Transition: $6p_{3/2} m_j~-~6s$} & \multicolumn{2}{c}{$|m_j|=1/2$} &
\multicolumn{2}{c}{$|m_j|=3/2$} \\ [0.2pc]
\hline
&$6p_{3/2} -   5d_{3/2}$ & 3.166(16) &   3614.09 &           &           &            &\\
&            &           & 	       & 3611.8(2) &  422(1)   &   3589(1)  &  422(1)\\
&$6p_{3/2} -   5d_{5/2} $ & 9.59(8)   &   3490.97 &  &&&\\
&            &           & 	       & 1910(6)   &  498(2)&&\\
&$6p_{3/2} -    7s $ & 6.47(3)   &   1469.89 &  &&&\\
&            &           & 	       & 932.4(8)  &  3197(50) &  940.2(1.7)&  2810(70)\\
&$6p_{3/2} -    6d_{3/2}$ &  2.10(5)  &   921.11  & &&&\\
&            &           &	       & 921.01(3) &  4088(10) &  920.18(6) &  4180(14)\\
&$6p_{3/2} -    6d_{5/2} $ &  6.15(14) &   917.48  &&&&\\
&            &           &	       & 887.95(10)&  -5600(100) & 883.4(2) & -1550(90)\\
&$6p_{3/2} -    6s  $&  6.324(7) &   852.35  &&&&\\
&$6p_{3/2} -    8s $ &  1.461    &   794.61  &&&&\\
&            &           &	       & 793.07(2) &  -2074(5) &  &\\
&$6p_{3/2} -    7d_{3/2} $&  0.976(9) &   698.54  & &&&\\
&            &           &	       & 698.524(2)&  -697(2)  &  698.346(4)&  -696(2)\\
&$6p_{3/2} -    7d_{5/2} $ &  2.89(3)  &   697.52  & &&&\\
&            &          &	       & 687.3(3)  &  -635(3)  &   684.1(5) &  -618(4)\\
&$6p_{3/2} -    9s $ &  0.770    &   658.83  &&&&\\
&            &         &	       & 657.05(9) &  -500(1)&&\\
&$6p_{3/2} -    8d_{3/2}$ &  0.607(8) &   621.93  & &&&\\
&            &         &	       & 621.924(2)&   -388(1) &  621.844(3)&   -388(1)\\ 
&$6p_{3/2} -    8d_{5/2}  $&  1.81(2)  &   621.48  & &&&\\
&            &         &	       &  615.5(8) &   -371(3) &     614(3) &   -367(8)\\
&$6p_{3/2} -   10s $ &  0.509    &   603.58  &&&&\\
&            &         &	       &  602.6(4) &   -339(1) &    &  \\
\end{tabular}   
\end{ruledtabular}
\end{table*}

\subsection{Rb}

We list the magic wavelengths $\lambda_{\rm{magic}}$ above 600 nm
  for the $5p_{1/2}-5s$ transition in Rb and the $6p_{1/2}-6s$ transition in Cs in Table \ref{tabRbCs1}. 
In this case, all Rb $5p_{1/2}-nl_j$ resonances have significant spacing allowing us
to determine the corresponding magic wavelengths. 
The magic wavelengths above 600 nm for the $5p_{3/2}-5s$ transition in Rb and $6p_{3/2}-6s$ transition in Cs
are grouped together in Table \ref{tabRbCs3}.
The transitions up to $5p-10s$
   and $5p-9d$
 are included in the main term calculation of the Rb $5p$ polarizabilities and the remainder is evaluated in the DF approximation.
The  $5p-5s$ matrix elements are taken from Ref.~\cite{volz}, and the $5p-4d$  E1 matrix elements 
are the recommended values derived from the Stark shift measurements \cite{milhun} in Ref.~\cite{KRb}. 
As we discussed in Section~\ref{pol}, the correlation correction is very large for the $5p-5d$ transitions; the DF
values for the $5p_{3/2}-5d_{5/2}$ transition is 0.5~a.u. while our final value is  2.0~a.u.  
  However, 
nearly entire correlation correction to this value comes from the single all-order term 
 which can be more accurately estimated by the scaling 
procedure described in Refs.~\cite{relsd,CC2,1}.
To evaluate the uncertainty of these values, we also conducted another calculation 
including the triple excitations relevant to the correction of the dominant 
correlation term (SDpT method), and repeated the scaling procedure for the SDpT calculation. We took the 
spread of the final values and the SDpT \textit{ab initio} and SDpT scaled values to be the 
uncertainty of the final numbers. Nevertheless, even  such an elaborate calculation still 
gives an estimated uncertainty of 4.5\%. 

We illustrate the $\lambda_{\rm{magic}}$ for the $5p_{3/2}-5s$ 
transition near 791 nm  in Fig. \ref{fig6}. 
 We note that this case is different from that of Na 
illustrated in Fig.~\ref{figna-3}, where both $\alpha_0+\alpha_2$ and 
$\alpha_0-\alpha_2$ curves for the $3p_{3/2}$ polarizability
cross the $3s$ polarizability curve at approximately the same polarizability values. 
In the Rb case near 791~nm,  $\alpha_0+\alpha_2$ and 
$\alpha_0-\alpha_2$ curves for the $5p_{3/2}$ polarizability
cross the $5s$ polarizability curve at 125~a.u. and
-6910~a.u., respectively. As a result, the $|M^{\prime}|=3$ curve on the  
 ac Stark shift plot for the transition between 
hyperfine sub levels shown in Fig.~\ref{fig7} is significantly split from the 
curves for the other sublevels. 

The magic wavelengths for the $5p_{3/2}-5s$ transition  between the fine-structure components 
of the $5p-nd_j$ levels are not listed owing to very small fine structures of these
levels. We note that crossings for all $m_j$  sublevels should be present between the fine-structure 
components of the $5p-nd_j$ lines. We illustrate such magic wavelengths for Cs, which 
has substantially larger $nd_j$ fine-structure splittings. 
	
		\begin{figure}[!htb]
  \includegraphics[scale=0.68]{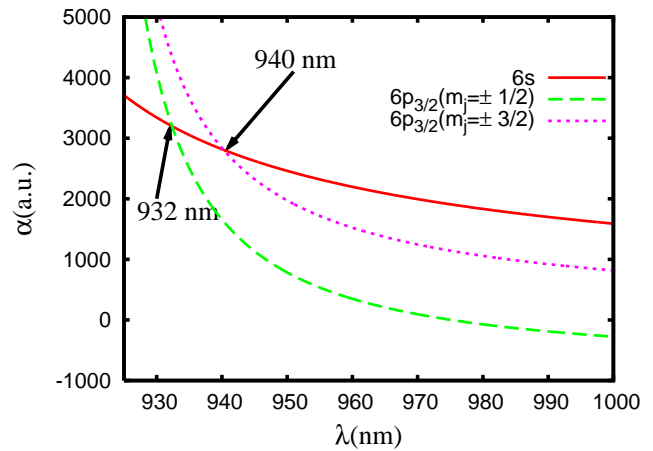}
  \caption{Polarizability of the $6p_{3/2}$ state and polarizability of  ground state of Cs as a function of 
wavelength. Magic wavelengths for the $6p_{3/2}-6s$ transition of Cs are found to be at 932 and 940 nm depending on the 
$m_j$ value.} 
  \label{fig8}     
	\end{figure}
	
	\begin{figure}[!htb]
  \includegraphics[scale=0.68]{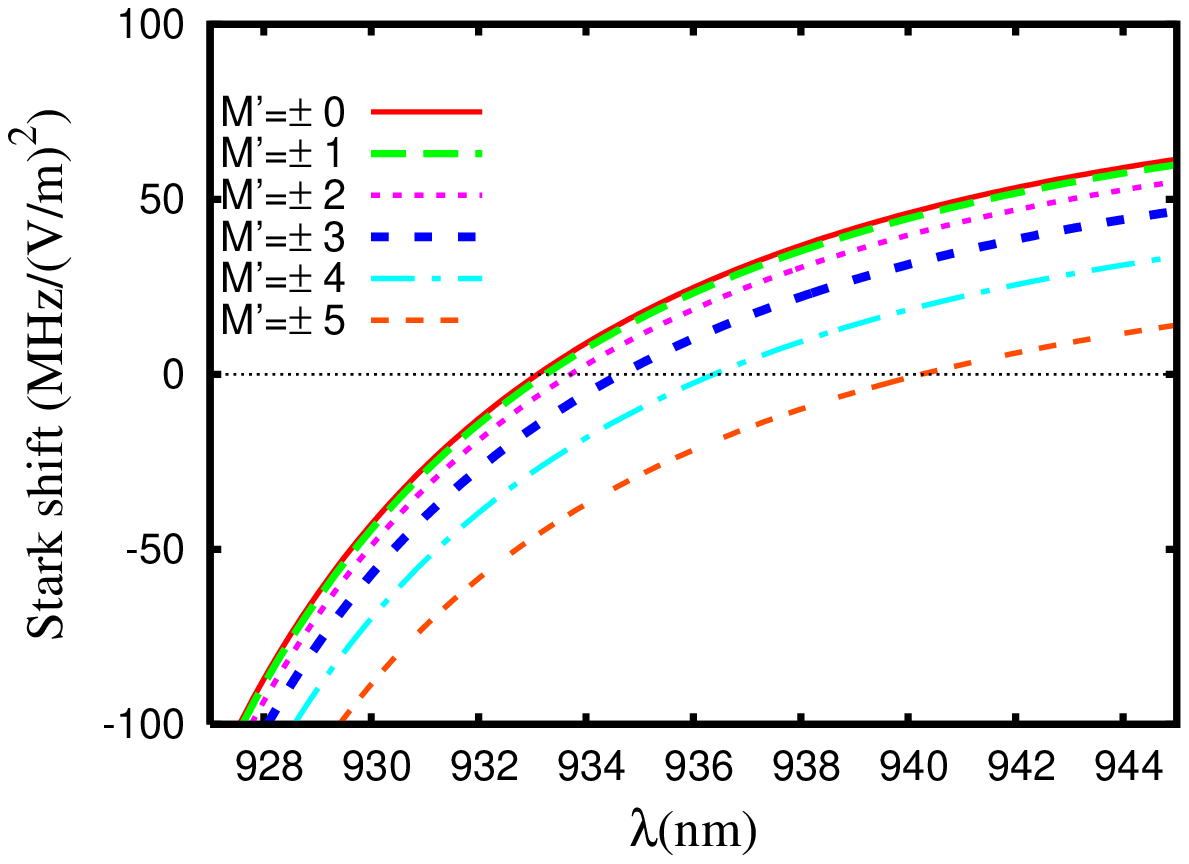}
  \caption{
      ac Stark shifts for the transition from $6p_{3/2} F'=5 M'$ sub levels to $6s FM$ sub levels in Cs as a
function of wavelength. The  electric field intensity is taken to be 1 MW/cm$^2$.
}
  \label{fig9}     
	\end{figure}

 \subsection{Cs}	 
Our results for Cs are listed in Tables~\ref{tabRbCs1} and \ref{tabRbCs3}.
  The values of the $6p-6s$ matrix elements are taken from \cite{rafac}, and
the values for $6p-7s$ transitions are taken from the  results compiled in \cite{relsd}
(derived from the $7s$ lifetime value). We derived the $6p_{1/2}-5d_{3/2}$
value from the experimental value of the D1 line Stark shift in Cs \cite{milhun2}
combined with the experimental ground state polarizability value from \cite{amini}. 
The procedure for deriving the matrix element values from the Stark shifts 
is described in Ref.~\cite{KRb}. We use the theoretical values of the ratios of the 
  $6p_{1/2}-5d_{3/2}$, $6p_{3/2}-5d_{3/2}$, and  $6p_{3/2}-5d_{5/2}$ 
  values from Ref.~\cite{1} to obtain the values for the $6p_{3/2}-5d_{3/2}$ and $6p_{3/2}-5d_{5/2}$
  matrix elements. We use the experimental energy levels from \cite{NIST,NIST2,SW}, and references
  therein to obtain the resonance wavelength values. The transitions up to $6p-12s$
   and $6p-9d$ are included into the main term calculation of the polarizabilities 
   and the remainder is evaluated in the DF approximation.

  We find that there are no magic wavelengths for the $6p_{1/2}-6s$ transition in between the 
  $6p_{1/2}-6s$, $6p_{1/2}-6d_{3/2}$, and  $6p_{1/2}-8s$ resonances whereas there are the magic wavelengths in between 
  the corresponding resonances in Rb. The difference between the Rb and Cs cases is  in the 
  placement of the $6p_{3/2}-6s$ resonance in Cs and $5p_{3/2}-5s$ resonance in Rb. In Rb,  $5p_{3/2}-5s$ resonance
  is at 780~nm and  follows the $5p_{1/2}-5s$ one. In Cs, the $6p_{3/2}-6s$ resonance
   is at 852~nm and is located in between the $6p_{1/2}-6d_{3/2}$ and $6p_{1/2}-8s$ resonances owing to much 
   larger $6p$ fine-structure splitting. As a result, there are no magic wavelengths in this range. 

Also unlike the Rb case, the magic wavelengths around 935~nm for the $6p_{3/2}-6s$ transition in Cs 
correspond to similar values of the polarizability  (about 3000~a.u.) for all sublevels as illustrated 
 in Fig.~\ref{fig8}. The nearest resonances to this magic wavelength are $6p_{3/2}-6d_j$ ones; therefore the 
 contributions from these transitions are dominant. To improve the accuracy of these values, we
  conducted a more accurate calculation for these 
 transitions following the $5p-5d$ Rb  calculation described in the previous subsection.
  As a result, we expect our values of the $6p-6d$ matrix elements to be more accurate
 than the one quoted in Ref.~\cite{1}.  Nevertheless, the uncertainties in the values of the corresponding 
 magic wavelengths are quite high because the $6s$ and $6p_{3/2}$ polarizability curves cross at very 
 small angles. As a result, even relatively small uncertainties in the values of the polarizabilities
 propagate into significant uncertainties in the values of the magic wavelengths.   
Our values for these magic wavelengths are in good agreement with previous 
studies~\cite{McKeever,kien}. 
The ac Stark shift of the $6p_{3/2}F'=5 M'$ to  $6s FM$ 
transition as a function of wavelength at the $925 - 945$~nm range is plotted in Fig. \ref{fig9}.	

\section{Conclusion}
We have calculated 
the ground $ns$ state and  $np$ state ac polarizabilities 
in Na, K, Rb, and Cs using the relativistic all-order method and evaluated the uncertainties of these values.  
The static polarizability values were found to be 
in excellent agreement with previous experimental and theoretical results. 
We have used our calculations
 to identify the magic wavelengths 
at which the ac polarizabilities of the alkali-metal atoms in the ground state are equal to the 
ac polarizabilities in the excited $np_j$ states 
 facilitating state-insensitive cooling and trapping.

\section{Acknowledgments}
We gratefully acknowledge helpful discussions with Fam Le Kien.
This work was performed under the sponsorship of the National Institute of Standards and 
Technology, U.S. Department of Commerce.

\end{document}